\documentclass[aps, 12pt,superscriptaddress, amsmath, amssymb, noshowpacs, preprint,preprintnumbers, nofootinbib]{revtex4}
\usepackage{amsmath}
\usepackage{amsfonts}
\usepackage{amssymb}
\usepackage{textcomp}
\usepackage{graphicx}
\usepackage{bm}
\usepackage{color}
\usepackage{verbatim}
\usepackage{bbm}

\usepackage{epsfig}
\usepackage{amsfonts,amsmath,amssymb}

\let\oldFootnote\footnote
\newcommand\nextToken\relax

\renewcommand\footnote[1]{%
    \oldFootnote{#1}\futurelet\nextToken\isFootnote}

\newcommand\isFootnote{%
    \ifx\footnote\nextToken\textsuperscript{,}\fi}

\def\beq{\begin{equation}}
\def\eeq{\end{equation}}
\def\beqn{\begin{eqnarray}}
\def\eeqn{\end{eqnarray}}
\newcommand{\be}{\begin{equation}}
\newcommand{\ee}{\end{equation}}
\newcommand{\ba}{\begin{eqnarray}}
\newcommand{\ea}{\end{eqnarray}}
\newcommand{\ben}{\begin{enumerate}}
\newcommand{\een}{\end{enumerate}}

\newcommand{\p}{\partial}

\newcommand{\la}{\langle}
\newcommand{\ra}{\rangle}
\newcommand{\lr}{\leftrightarrow}

\newcommand{\rar}{\rightarrow}

\begin{document}

\preprint{NORDITA-2015-29}

\title{Baryon and chiral symmetry breaking in holographic QCD}

\author{Alexander Gorsky}
\affiliation{Institute of Information Transmission Problems, \\
B. Karetnyi 19, Moscow, Russia}
\affiliation{Moscow Institute of Physics and Technology, \\
Dolgoprudny 141700, Russia}

\author{Sven Bjarke Gudnason}

\affiliation{Institute of Modern Physics, Chinese Academy of Sciences, 
NanChangLu 509, Lanzhou 730000, China} 

\affiliation{Nordita, 
KTH Royal Institute of Technology and Stockholm University \\
Roslagstullsbacken 23, SE-106 91 Stockholm, Sweden. }

\author{Alexander Krikun}

\affiliation{Nordita, 
KTH Royal Institute of Technology and Stockholm University \\
Roslagstullsbacken 23, SE-106 91 Stockholm, Sweden. }

\affiliation{Institute for Theoretical and Experimental Physics (ITEP), \\ 
B. Cheryomushkinskaya 25, 117218 Moscow, Russia}

\begin{abstract}
We study the relationship between chiral symmetry breaking and baryons
in holographic QCD. We construct a soliton with unit baryon charge in
the presence of a nonzero mean value of the scalar bifundamental field,
which is dual to the chiral condensate. We obtain a relation between
the chiral condensate and the mass of the baryon and find in a clear-cut
way that at large values of the condensate the holographic soliton is no longer
located on the IR wall. Instead it is split into two halves, which are
symmetrically located on the left and right flavor branes. On the
other hand we find that the local value of the quark condensate is
suppressed in the core of the soliton, which is evidence for
a partial chiral symmetry restoration inside the baryon.  
\end{abstract}

\maketitle
 
\section{Introduction}

Baryon physics in holographic setups has been addressed since the time
the first holographic QCD models have been developed. There is a
wealth of works related to the interpretation of the baryon as a
holographic instanton \cite{Hata:2007mb, Bolognesi:2013nja, Rozali:2013fna}, the QCD currents sourced by the baryon
\cite{Hashimoto:2008zw, Hong:2007dq, Kim:2008pw} and even high-density
baryonic matter \cite{Kaplunovsky:2012gb, Dymarsky:2010ci,
Rho:2009ym}. Unfortunately though, the studies of the interplay between baryon physics and physics of chiral symmetry breaking is undeservedly rare: Ref.\cite{Domenech:2010aq} is the only example that we are aware of. But from the other hand these relationships remain one of the longstanding questions in QCD \cite{ioffe} and should be definitely addressed with the new tools provided by the holographic duality, which proved to be useful in describing both chiral symmetry breaking and baryons in QCD.

Usually the AdS/QCD model, which is used for the
studies of baryons, is (a modification of) the Sakai-Sugimoto model (SS)
\cite{Sakai:2004cn, Sakai:2005yt}. This model is based on the system
of $N_f$ $D8-\overline{D8}$ branes, which are embedded in the curved
hyperbolic space produced by a stack of $N_c$ $D4$ branes. One of the 
dimensions, $\tau$, is compact and serves for breaking supersymmetry. The
compactification radius of $\tau$ vanishes at some point and thus the
background space has the shape of a cigar \cite{Witten:1998zw}. The
location of the tip of  
the cigar plays the role of the $\Lambda_{QCD}$ scale. From the
holographic point of view, the tip serves as a boundary beyond which
nothing can propagate as the space practically ends there. The
$D8-\overline{D8}$ branes embedded in this background are located at the
opposite (or almost opposite \cite{Seki:2008mu}) points of the
$\tau$-circle and consequently they merge together when the circle
collapses. The gauge field, $\mathcal{A}$, which is a massless mode of
the open string with two ends connected to the $D8$ branes, belongs to
the adjoint representation of the flavor group $U(N_f)$ and is dual to
the quark current operator of QCD. One can introduce the coordinate
$U$ along the branes with the origin on the tip of the cigar, which
goes from~$-\infty$ on the $D8$ brane to $+\infty$ on the
$\overline{D8}$ brane. 
The values of the gauge field $\mathcal{A}$ at $-\infty$ may be
thought of as being coupled to the left quark current 
$J^{a L}_\mu=\bar{q} \frac{1 + \gamma_5}{2} \gamma_\mu t^a q$ 
while the values at $\infty$ are coupled to the right current  
$J^{a R}_\mu=\bar{q} \frac{1 - \gamma_5}{2} \gamma_\mu t^a q$. 
Accordingly one can call the $D8$ branes Left and the $\overline{D8}$
branes Right.  

The baryon in the Sakai-Sugimoto model is an instanton of the field
$\mathcal{A}$ \cite{Hata:2007mb}. The baryon charge equals the topological
charge of the instanton because the baryon current is sourced in the
Chern-Simons (CS) term of the action by a topological charge
density. One can find that the instanton in the SS model is well
approximated by the usual flat BPST instanton. This holds because the
size of the instanton is parametrically small; the center of the
solution is located on the tip of the cigar and one can neglect the
curvature of the metric at the scale of the solution radius. The
vanishing difference between spatial ($x_i, i=1,\dots,3$) and
holographic ($U$) parts of the metric leads to an approximate $SO(4)$
symmetry group of the space \cite{Bolognesi:2013nja}. In this setting
it is very convenient to use the usual radial ansatz of the BPST
instanton with the radial coordinate $\rho^2 = U^2 + \Sigma x_i^2$. The
solution is then located at $\rho=0$. 

For our purposes, though, the Sakai-Sugimoto model is not entirely
satisfactory. The obstacle is that it is not at all easy to describe
chiral symmetry breaking and the associated chiral condensate in this
setup. This problem was analyzed in Refs.~\cite{Bergman:2007pm, Dhar:2008um}
and it was shown that the holographic field dual to the scalar quark
current $\bar{q}q$ is a tachyonic mode of the string stretched between
the $D8$ and the $\overline{D8}$ branes. The tachyonic field condenses
on the tip of the cigar, thus providing a nonzero vacuum expectation
value (VEV) of $\bar{q}q$, the chiral condensate. There are, however,
several complications in this setup. The construction of the tachyonic
DBI action is ambiguous, the backreaction of the condensed open string
on the geometry of space should be taken into account and so on. On top
of that, when constructing an effective 5-dimensional holographic model
for mesons it is impossible to treat the open string as a local field. 

The chiral symmetry breaking is relatively easy to implement in the
other, ``bottom-up'', approach of holographic QCD. Instead of
constructing the brane system which would reproduce the effective
theory of mesons, one builds the model by including the fields, dual
to the QCD operators, in the 5-dimensional AdS background
\cite{Erlich:2005qh, Karch:2006pv}. In the ``hard-wall'' (HW) model
\cite{Erlich:2005qh} the space ends on the wall located at a finite
value of the 
holographic coordinate $z_m$. This wall (we will call it the IR wall)
serves the same purpose as the tip of the cigar in the SS model by
breaking conformal symmetry and providing the scale
$\Lambda_{QCD}$. There are \textbf{two} gauge fields $\bm{L}$ and
$\bm{R}$, which are dual to left and right quark currents. The
field, which is dual to the scalar quark operator $\bar{q}q$, is just
a bifundamental scalar $\bm{X}$. Its vacuum profile is described
by the two parameters 
\begin{align}
\label{vac_X}
\bm{X}_0 = \frac{\mathbbm{1}}{2} \big(m z + \sigma z^3 \big),
\end{align}
which by the holographic dictionary \cite{gubser-klebanov} are related
to the quark mass $m$ and condensate (see Refs.~\cite{Krikun:2008tf,
  Gorsky:2009ma, Cherman:2008eh} and the derivation in Sec.~\ref{sec:form}) 
\begin{equation}
\label{sigma}
\la \bar{q} q \ra = \frac{N_c}{2 \pi^2} \sigma. 
\end{equation}
In the model there is no internal mechanism, which would fix the value
of the chiral condensate. Instead one obtains a nonzero value of
$\sigma$ by imposing a Dirichlet boundary condition on $\bm{X}$
at the IR wall 
\begin{equation}
\label{bc_X} 
\bm{X} \Big|_{z=z_m} =\frac{\mathbbm{1}}{2} \big( m z_m + \sigma z_m^3
\big) = {\rm const}.
\end{equation}
One can obtain the correlation functions of various operators in the
setup with nonzero profile of the bifundamental scalar and study the
dependence of various observables on the chiral condensate, which is
just a parameter of the model. Unfortunately, the treatment of the baryon
in this setup is not as straightforward as it is in the SS model. There
are two gauge fields, which should realize the topologically
nontrivial solution, and on top of that, there is a scalar with a nonzero
VEV, which interacts with both of them. One should also pay special
attention to the boundary conditions at the IR boundary, as they are
arbitrary in the construction and should be additionally fixed in the
model. Concerning the baryon, the question of IR boundary condition becomes
substantially important because the corresponding topological soliton
falls on the IR wall, a phenomenon similar to the localization of
the SS-instanton on the tip of the cigar. The baryon physics in HW
model without a chiral condensate was addressed  
in Ref.~\cite{Pomarol:2008aa}.

The bifundamental scalar was included in Ref.~\cite{Domenech:2010aq},
however the dependence of the baryon mass on the 
condensate has not been investigated. Motivated by the formula derived
using the QCD sum rules in Ref.~\cite{ioffe} two of us  
have analyzed the baryon-mass origin in the framework of the hard wall
model \cite{gk, gkk} and found that at 
large values of the chiral condensate, there are clear indications
that it dominates. However the accuracy of the numerical calculations
was not high enough to make precise claims. In this paper we perform
a detailed analysis of the different aspects of the impact of the chiral
symmetry breaking on the baryon solution. We will clearly see the
internal structure of the solution and two regimes in the
dependence of the baryon mass on the chiral condensate.  

The paper is organized as follows. In Sec.~\ref{sec:model} we
introduce our setup, which we try to keep sufficiently general and
applicable to a wide range of holographic QCD models. We study the
qualitative structure of the baryon solution in the presence of the chiral
scalar field in Sec.~\ref{sec:structure}. The quantitative support for
this treatment is given by the numerical solution in
Sec.~\ref{sec:numerics}. Secs.~\ref{sec:mass} and \ref{sec:form} are
devoted to the calculation of the baryon mass and local mean values of
the QCD currents in presence of the baryon. We conclude in
Sec.~\ref{sec:conclusion}. The analysis of the precision of our
numerics is outlined in the Appendix.

\section{\label{sec:model} The model}

\begin{figure}[ht]
 \includegraphics{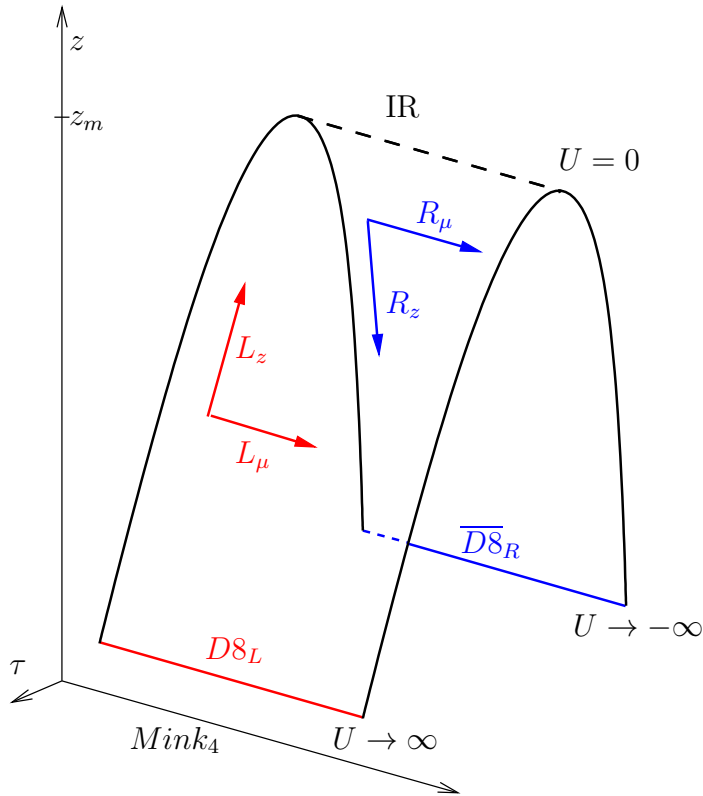}
 \caption{\label{folding} Scheme of the ``folding trick'' explained in the text.}
\end{figure}

It is actually possible to make a qualitative connection between the
seemingly different setups of the Sakai-Sugimoto and the ``hard wall'' 
models. In order to build up our intuition we use the following
``folding trick'': Consider the cigar-bended $D8$ brane of the SS
model. There is a single gauge field $\mathcal{A}$ on this brane, but
let us call it $\bm{L}$ when it is considered on the Left brane
and $\bm{R}$ on the Right brane (see Fig.\,\ref{folding}). On the
tip of the cigar $\bm{L} = \bm{R}$, of course. Next, let us
assume trivial dynamics along the $\tau$ direction, which will allow
us to neglect the separation of the branes along $\tau$ altogether,
effectively ``folding'' the cigar-shaped brane into a single
sheet. More specifically, we cut the brane along the tip or $U=0$
line. On one side we have the $\bm{L}$ field and $U \in [0; +\infty)$
and on the other side the $\bm{R}$ field and $U \in (-\infty; 0]$. In
order to get consistent coordinates, we perform a $U \rar - U$
transformation on the Right brane, which also means $\bm{R}_U
\rar -\bm{R}_{U}$. It is convenient to introduce the inverse
holographic coordinate $z = 1/(U + z_m^{-1})$. Then the tip of the
cigar ($U=0$) corresponds to the maximum $z=z_m$ and the boundary
values are assigned at $z=0$. At the end of the day we have two gauge
fields propagating on a strip $z \in (0,z_m]$, which is qualitatively 
the same as in the HW model. The important additional information, that
we get here, is the boundary conditions on the gauge fields at the IR
boundary. Because we know that $\bm{L}$ and $\bm{R}$ are
the same field there, we get 
\begin{align}
\label{bc_IR}
\mathrm{IR:}\qquad  \bm{L}_z + \bm{R}_z = 0, \qquad \bm{L}_\mu - \bm{R}_\mu = 0.
\end{align}
In this picture the tachyonic string, which was responsible for the
chiral condensate in the SS model, collapses to the local bifundamental
tachyonic scalar $\bm{X}$. The SS baryon, which was originally
located at $U=0$, is now split into two halves, one composed of
$\bm{L}$ and the other of $\bm{R}$, which are folded
together. Thus we expect that in the HW model, the baryon would appear as
one half of the instanton lying on the IR wall.

As we see, in any case the holographic QCD model with two quark
flavors $N_f=2$ can be described by dynamics of left ($\bm{L}$) and
right ($\bm{R}$) gauge fields of the group $U(2)_L \times U(2)_R$ and
a bifundamental tachyonic scalar $\bm{X}$ in 5-dimensional curved
space. We use Hermitian gauge-group generators and treat the gauge
connection as Hermitian matrices\footnote{The corresponding definition of
the gauge field strength is $\bm{F}_{\mu \nu} = \p_\mu \bm{A}_\nu -
\p_\nu \bm{A}_\mu - i [\bm{A}_\mu, \bm{A}_\nu]$}: 
\begin{gather}
\bm{L} = \hat{L} \ \frac{\mathbbm{1}}{2} + L = \hat{L} \ \frac{\mathbbm{1}}{2} + L^a t^a, \qquad  a=1,\dots,3,  \\
\notag
t^a = \frac{\sigma^a}{2}, \qquad \mathrm{tr}(t^a t^b) = \frac{1}{2} \delta^{ab}, \qquad  [t^a, t^b]= i \epsilon^{abc} t^c.
\end{gather}
where $\sigma^a$ are the Pauli matrices. 
The background space can be described by the mostly-minus metric of
the general form\footnote{Note the peculiar numbering of the
  coordinates: we never use $x_4$ as it usually stands for 
  Euclidean time and our treatment is Minkowskian. For the holographic
  coordinate we use $x_5$ instead to avoid possible confusion. } 
\begin{equation}
ds^2 = h(z)^{2} dt^2 - h(z)^{2} dx_i^2 - k(z)^{2} dz^2, \qquad 
t\equiv x_0, \ z\equiv x_5, \ i=1, \dots, 3.
\end{equation}
For instance, in the ``hard-wall'' model \cite{Erlich:2005qh} 
$k(z)=h(z)= L/z$; in Sakai-Sugimoto model \cite{Sakai:2004cn} 
$h(z)= (L/z)^{3/4}, \ k(z) = (L/z)^{1/4}
(1-\frac{z^3}{z_m^3})^{-1/2}$. 
$L$ is a curvature scale and the holographic coordinate $z$ is bounded
from above: $z \in (0; z_m]$. The Yang-Mills action is 
\begin{align}
\label{action.gen}
S_{YM} = \int d^3 x dt dz \bigg\{ - \frac{1}{4 g_5^2}  & \mathrm{Tr}\Big\la  k(z) {\bm{F}_L}_{ij}^2 - 2 k(z) {\bm{F}_L}_{0i}^2 + 2  \frac{h(z)^2}{k(z)} \left({\bm{F}_L}_{5i}^2 -{\bm{F}_L}_{50}^2  \right)  + (\bm{L} \lr \bm{R})\Big\ra \\
\notag
 + g_X^2  & \mathrm{Tr}\Big \la h(z)^2 \left( D_0 \bm{X}^2  - D_i \bm{X}^2 \right) -  \frac{h(z)^4}{k(z)} (D_5 \bm{X})^2 - h(z)^4 k(z) m_X^2 |\bm{X}|^2 \Big\ra \bigg\},
\end{align}
where the covariant derivative is $D_* \bm{X} = \p_* \bm{X} - i
\bm{L}_* \bm{X} + i \bm{X} \bm{R}_*$ and the gauge coupling $g_5$ and
normalization of the scalar $g_X$ are to be fixed by phenomenology
\cite{Erlich:2005qh, Krikun:2008tf, Gorsky:2009ma, Cherman:2008eh} or
a specific top-down construction \cite{Sakai:2004cn,
  Bergman:2007pm}\footnote{Importantly, in the Sakai-Sugimoto model, $g_5$ 
  is inversely proportional to the 't Hooft coupling of the dual field theory, which is
  assumed to be large.}. The mass of the scalar $m_X$ is defined by
the dimension of the corresponding operator $q^\alpha \bar{q}^\beta$
and equals $m_X^2 = - \frac{3}{L^2}$. 

The Chern-Simons term is present as well
\begin{equation}
\label{CS}
S_{CS} = \frac{N_c}{24 \pi^2} \int \frac{3}{2} \ \left \{ \hat{L} \wedge \mathrm{Tr} \left\la F_L \wedge F_L \right\ra - (\bm{L} \lr \bm{R}) \right\} + \dots.
\end{equation}
One can immediately spot that the non-Abelian part of the gauge field,
which has nonzero topological charge in the 4-dimensional spacelike
slice, provides a source for the temporal component of the Abelian
field, which is dual to the baryon current. Thus we can identify the
baryon charge of a given field configuration \cite{Hata:2007mb,
  Pomarol:2008aa}   
\begin{equation}
\label{baryon_charge}
Q_B =\frac{1}{16 \pi^2} \int d^3 x dz \; \epsilon^{m n l k} \left({F_L^a}^{m n} {F_L^a}^{l k} - {F_R^a}^{m n} {F_R^a}^{l k} \right), \qquad m = 1,\dots, 3, 5.
\end{equation}

In order to construct the solution in the 4D spacelike $(x_i, z)$-plane we
adopt the cylindrical ansatz proposed in Refs.~\cite{Witten_inst,
  Pomarol:2008aa, Domenech:2010aq}, taking $r=\sqrt{x_i^2}$ and $z$ as
the cylindrical coordinates. The 4D gauge potentials are parametrized
by all possible tensor structures which link the spatial and $SU(2)$
group indices. It is convenient to work with the vector and axial
combinations, defined as 
\begin{equation}
\bm{L}=\bm{V} + \bm{A}, \qquad  \bm{R}=\bm{V} - \bm{A}.
\end{equation}
In order to preserve 3D parity, we choose the P-odd and P-even tensors
for the spatial components of $V_i = (L_i+R_i)/2$ and
$A_i=(L_i-R_i)/2$, respectively. Similarly, the time and $z$ components
should be parity even for $V$ and odd for $A$. Hence we obtain 
the following ansatz for the gauge fields 
\begin{align}
\label{ansatz}
V_j^a &= -\frac{1 +  \eta_2 (r,z)}{r} \epsilon_{jak} \frac{x_k}{r}, & V_5^a &= 0, & \hat{V}_0 &= v(r,z),\\
\notag
A_j^a &=  \frac{\eta_1 (r,z)}{r} \left( \delta_{ja} - \frac{x_j x_a}{r^2} \right) + A_r(r,z)  \frac{x_j x_a}{r^2}, & A_5^a &= A_z(r,z) \frac{x_a}{r}, & \hat{A}_0 &= 0.
\end{align}
For the scalar we have
\begin{equation}
X = \chi_1(r,z) \frac{\mathbbm{1}}{2} + i \chi_2(r,z) \frac{t^a x^a}{r}. \\
\end{equation}
The action (\ref{action.gen}) with this ansatz is $S= \int dt dr dz \big( \mathcal{L}_{YM} + \mathcal{L}_{U1} + \mathcal{L}_{CS} \big)$:
\begin{align}
\label{SYM}
\mathcal{L}_{YM} &=  -\frac{2 \pi}{g_5^2} \bigg\{ 
r^2 \frac{h(z)^2}{k(z)}  \big[ \p_z A_r - \p_r A_z \big]^2   \\
\label{kin_eta}
&\phantom{=  -\frac{2 \pi}{g_5^2} \bigg\{\ }
+ \frac{2}{k(z)} \Big[ k(z)^2 \mathcal{D}_r \eta_\alpha^2 
+ h(z)^2
 \mathcal{D}_z \eta_\alpha^2  \Big] \bigg \}\\
 \label{kin_chi}
&- 2 \pi g_X^2  \bigg\{r^2  \frac{h(z)^2}{k(z)}  \Big[ k(z)^2 \mathcal{D}_r \chi_\alpha^2  
+ h(z)^2   \mathcal{D}_z \chi_\alpha^2 + m_X^2 h(z)^2 k(z)^2 \chi_\alpha^2 \Big]   \\
\label{potentials}
&\phantom{=  -\frac{2 \pi}{g_5^2} \bigg\{\ }
+ 2  \ \frac{h(z)^4}{k(z)}  \big[\chi_2 \eta_2 + \chi_1 \eta_1 \big]^2 
+ \frac{1}{g_X^2 g_5^2} \frac{k(z)}{r^2}  \left(1  - \eta_\alpha^2\right)^2 \bigg\}, \quad \alpha=1,2,\\
\label{SU1}
\mathcal{L}_{U1} &=  \left(\frac{4 \pi}{2 g_5^2}\right)  
\frac{r^2}{k(z)} \Big[ h(z)^2 (\p_z v)^2 + k(z)^2 (\p_r v)^2 \Big],\\
\label{SCS}
 \mathcal{L}_{CS} &=  \left( \frac{N_c}{2 \pi} \right) \ v \ \Big[
 2 \epsilon^{\alpha \beta} \mathcal{D}_z \eta_\alpha \mathcal{D}_r \eta_\beta 
  + 
 \big(\p_z A_r - \p_r A_z \big) \left(1  - \eta_\alpha^2\right) \Big],
\end{align}
where the Abelian covariant derivative is 
$\mathcal{D}_* \eta_\alpha = \p_* \eta_\alpha + A_* \epsilon^{\alpha \beta} \eta_\beta$, and similarly for $\chi_\alpha$.
We see that the problem boils down to the 2D Abelian Higgs model with
two complex scalars $\eta$ and $\chi$, which interact with each other,
plus some potentials as well as interactions with the real scalar $v$,
which has the opposite sign of the kinetic term. One can expect to
have vortices as topologically nontrivial solutions in this model
\cite{Witten_inst}.

\section{\label{sec:structure} Structure of the solution}

In order to study the structure of the soliton it is useful to
introduce the phases of the complex scalars 
\begin{align}
\label{phases}
\eta_1 + i \eta_2 = \eta e^{i \theta}, \qquad 
\chi_1 + i \chi_2 = \chi e^{i \gamma}.
\end{align}
First we note that the baryon charge (\ref{baryon_charge}) can be rewritten in the form
\begin{align}
Q_B =  \frac{1}{\pi} \int dr dz \Big[ \p_r(\mathcal{D}_z \eta_\alpha \eta_\beta \epsilon^{\alpha \beta}) - \p_z(\mathcal{D}_r \eta_\alpha \eta_\beta \epsilon^{\alpha \beta}) + \big(\p_z A_r - \p_r A_z \big)
  \Big].
\end{align}
Once we demand that, in a particular gauge, the first two total
derivative terms vanish\footnote{The boundary terms at $r\rar0$, 
  $r\rar \infty$ and $z\rar0$ vanish because the finite energy condition
  forces $\mathcal{D}_{z,r} \eta_\alpha =0$, while the term at $z=z_m$
  vanishes because of the boundary condition $\eta_1 \big|_{z_m}=0$. See
  details below.}, we are left with the standard topological charge of
Abelian Higgs model. Assuming the pure gauge condition 
$A_i = \p_i \theta + const$ at the boundaries, we realize that the
topological charge equals the winding of the phase $\theta$ around the
boundary of the spatial patch. 
\begin{align}
Q_B = \frac{1}{\pi} \int dr dz  \big(\p_z \p_r \theta - \p_r \p_z \theta \big).
\end{align}
One should note though that the baryon, which is a solution with
$Q_B=1$, should have the phase $\theta$ winding by $\pi$. Hence the
baryon in our model is a \textbf{half-vortex}.\footnote{This result is
  consistent with our general intuition of ``folding'' the
  Sakai-Sugimoto model, where the baryon is a full vortex, which lies
  on the tip of the cigar.}. The only way a half-vortex can be realized
as a smooth solution is by having its core located on one of the
boundaries. There is only one boundary of the 
$[0,\infty)_r \times (0,z_m]_z$ patch which is suitable for
this. Indeed, in the core of the vortex the modulus $\eta$ should
necessarily vanish. The second potential term in Eq.~(\ref{potentials})
does not allow $\eta$ to deviate from 1 either at $r=0$ or at $z=0$
where $k(z)$ is singular. The core cannot be located at infinity as
well, because this will require a nonzero derivative $\p_r \eta$ at
$r\rar\infty$ and lead to a divergence in energy due to the  
$\eta$ kinetic term (\ref{kin_eta}). Thus we see that the core of the
$\eta$ half-vortex should be located \textbf{on the IR wall} $z=z_m$,
which corresponds to the tip of the cigar in the Sakai-Sugimoto
picture. 

At the next step we should consider the interaction of the scalar
fields $\eta$ and $\chi$. It is governed by the first term in
Eq.~(\ref{potentials}), which conveniently can be represented in terms of
the phases (\ref{phases}):
\begin{align}
\label{phase_lock}
\frac{h(z)^4}{k(z)} \big[\chi_2 \eta_2 + \chi_1 \eta_1 \big]^2 = \frac{h(z)^4}{k(z)}  \chi^2 \eta^2 \cos(\theta - \gamma)^2.
\end{align}
We see that this interaction effectively ``locks'' the phases $\theta$
and $\gamma$, requiring $\theta-\gamma = \frac{\pi}{2} + \pi
n$. Therefore, if $\theta$ has a winding $\pi$, the same holds true for
$\gamma$ and the scalar $\chi$ realizes a \textbf{half-vortex} as
  well. The location of the core of this vortex is different
though. At infinity we need to connect the baryon solution with the
vacuum, where the profile of the $\bm{X}$ field is given by the
quark mass and condensate (\ref{vac_X}), thus the modulus $\chi$
cannot vanish. The core cannot be located at $z=0$ as well, because 
according the holographic dictionary the boundary value of $\chi$ is
defined by the source of the corresponding operator, which is the
quark mass. On the IR wall, $\chi$ is fixed by the boundary condition
(\ref{bc_X}) which defines the quark condensate\footnote{We note here,
  that this condition is absent in the Sakai-Sugimoto model. Instead one  
  should consider the behavior of the tachyonic open string near the tip
  of the cigar \cite{Bergman:2007pm}.}. At the end of the day, the only
remaining place for the core of the $\chi$ half-vortex is \textbf{the
  center of the soliton} $r=0$. 

\begin{figure}[ht]
 \includegraphics{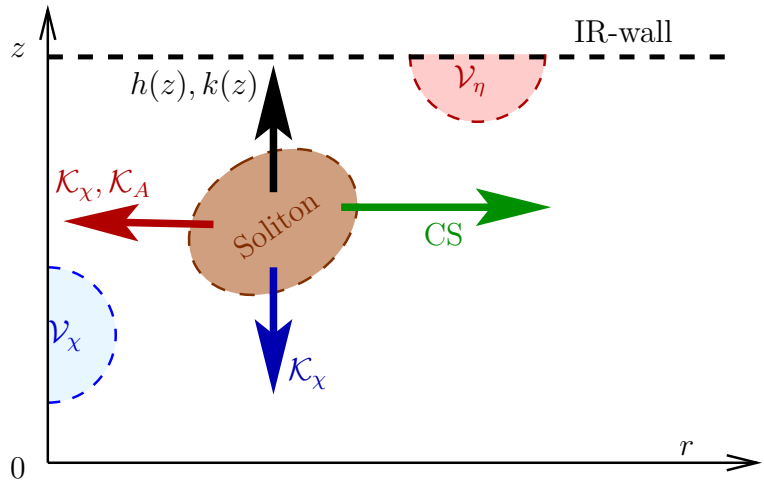}
 \caption{\label{scheme} Schematic structure of the
   soliton. $\mathcal{V}_\eta, \mathcal{V}_\chi$ -- half-vortices of
   $\eta$ and $\chi$ scalars. Arrows show the effective forces
   produced by the metric ($h(z),k(z)$), kinetic terms of $\chi$ and
   $A_\mu$ ($\mathcal{K}_\chi, \mathcal{K}_A$) and self interaction
   with the Abelian field in the CS-term. }
\end{figure}

At this point we see that the holographic baryon is a soliton
consisting of two interacting half-vortices of two scalar fields. One
is located on the IR wall and another at the center of the soliton
(see Fig.~\ref{scheme}). Now it is important to consider the energy of
the solution and ensure that there exists a mechanism which stabilizes
its size, preventing the soliton from collapsing.  

It was pointed out in Ref.~\cite{Hata:2007mb} that due to the metric
factors $h(z), k(z)$ the energy of the soliton is roughly inversely
proportional to its $z$ position. Hence the solution tends to lie at
the largest possible $z = z_m$. In other words, the metric provides a
force, which \textbf{pulls the solution to the IR wall}.  

On the other hand, due to the IR boundary condition on $\bm{X}$,
the core of the $\chi$ vortex cannot approach IR
wall too close. This would lead to a large gradient of $\chi$, which 
evolves from zero in the core to $\chi \sim \sigma$ on the wall. Hence
the kinetic term of $\chi$ (\ref{kin_chi}) and IR boundary condition (\ref{bc_X})  provide a counterforce,
which \textbf{repels the solution from the wall}.  

One can see though that this counterforce is proportional to the
radius of the solution squared. Therefore it becomes negligible when
the size of the soliton is small. Moreover, the whole $\chi$ energy
term (\ref{kin_chi}) is proportional to $r^2$, so it contributes (in
addition to the gauge field kinetic term (\ref{SYM})) to the pressure,
which would \textbf{shrink} the solution to zero size.\footnote{It was
  noted in Ref.~\cite{Domenech:2010aq} that the solution can be stable even
  in the absence of the Chern-Simons term. Our treatment shows that this
  should not be the case. Indeed, when the solution shrinks the
  contribution from $\chi$ weakens and cannot prevent further
  shrinking. } 

It is known \cite{Hata:2007mb, Pomarol:2008aa} that the important
counterforce to this shrinking, which stabilizes the radius of the
solution, is provided by the interaction of the soliton with the
Abelian gauge field via the Chern-Simons term (\ref{SCS}). The smaller the
radius of the solution becomes, the higher the density of
topological charge is, which sources the temporal component of the vector
Abelian field, and the larger its gradient is. This results in the
effective \textbf{internal pressure} which fixes the size of the
soliton (see Fig.\,\ref{scheme}).

This treatment shows an interesting effect, which is produced by
the chiral scalar field in the ``hard-wall'' model. Once the soliton
radius is fixed by the CS term, the IR boundary condition (\ref{bc_X})
together with the $\chi$ kinetic term (\ref{kin_chi}) produce the
force which repels the soliton from the IR wall. The larger the chiral
condensate is, the stronger the repulsion is. At large enough values of
$\sigma$, this repulsion will define the equilibrium $z$ coordinate of
the soliton center and therefore, will directly affect its
energy. Hence one should expect a substantial dependence of the baryon
mass on the chiral condensate. In the Sakai-Sugimoto model, though, the
radius of the soliton is inversely proportional to the 't Hooft coupling
and is therefore parametrically small \cite{Hata:2007mb,
  Bolognesi:2013nja}. This renders the repulsive force parametrically
small as well and consequently it is questionable whether in the Sakai-Sugimoto model  any impact of the chiral condensate on the baryon mass can be observed.

\section{\label{sec:numerics}Numerical solution}

In order to confirm the expectations outlined above
we construct the soliton solution numerically. For the numerical
simulation we choose the ``hard-wall'' model, so the metric is
unperturbed AdS$_5$: $h(z)=k(z)=L/z$, and the scalar asymptotic
profile is given by Eq.~(\ref{vac_X}). The couplings in the model have the
values \cite{Erlich:2005qh, Cherman:2008eh, Gorsky:2009ma}
\begin{align}
g_5^2 = \frac{12 \pi^2}{N_c}, \qquad g_X^2 = \frac{3}{g_5^2}.
\end{align}
The IR wall is located at $z_m = (323 \,\mathrm{MeV})^{-1}$ which is
fixed by the mass of the $\rho$-meson \cite{Erlich:2005qh}. The parameter
$\sigma$ is related to the quark condensate as in Eq.~(\ref{sigma}). In
what follows, unless explicitly stated otherwise, we will measure all
the dimensional quantities in terms of $z_m$ and rescale the curvature
radius $L=1$.

Because we are dealing with a gauge field theory, first of all we
need to fix the gauge. It is useful to adopt the Lorenz gauge 
$\p_z A_z + \p_r A_r = 0$. 
The gauge fixing condition is enforced by adding the Lagrange term
to the action  
\begin{equation}
\label{gauge_fix}
S_\lambda = \bm{\lambda} \left(-\frac{4 \pi}{2 g_5^2} \right) r^2 \frac{h(z)^2}{k(z)} \left[\p_z A_z + \p_r A_r \right]^2. 
\end{equation}
With this choice of gauge fixing, the equations for $A_z$ and $A_r$
become elliptic and the boundary value problem has unique solution. The resulting physical solution should not 
depend on the value of $\bm{\lambda}$. 

Next we need to choose the boundary conditions consistently with the
given gauge. In the $\eta$ half-vortex the phase $\theta$ changes by
$\pi$ along the boundary of space. We can chose the gauge in such
a way that the total winding occurs along the 
$r \rar \infty$ boundary, where $\theta$ changes linearly 
$\theta \big|_{r \rar\infty} = - \frac{\pi}{2} + \pi \frac{z}{z_m}$ 
and it is constant everywhere except the jump by $\pi$ in the core of
the $\eta$ half-vortex on the IR boundary. This choice of $\theta$
ensures us 
that $\eta_2 = -1$ at $z=0$, hence there are no source terms for the
axial current, $\eta_2=-1$ at infinity, hence the currents decay
there, and $\eta_1=0$ on the IR wall which is required by the IR
condition (\ref{bc_IR}). Due to the interaction between the scalars $\chi$ and
$\eta$ (\ref{phase_lock}), the phase difference $(\theta -
\gamma)$ should be either $-\frac{\pi}{2}$ or $\frac{\pi}{2}$ on all
the boundaries. Therefore the phase $\gamma$ winds linearly along 
the $r \rar \infty$  
boundary as well: $\gamma \big|_{x=1} = \pi \frac{z}{z_m}$. It is
constant along the $z=0$ and $z=z_m$ boundaries, having a $\pi$ jump in
the core of the $\chi$ half-vortex at $r=0$. One can check that this
choice of phase guarantees the constant value of the quark mass $m$
at $z=0$. 

The pure gauge condition at $r\rar \infty$ will then fix $A_z$ to be a
constant: $A_z\big|_{r \rar \infty} = \p_z \theta = \p_z \gamma =
z_m^{-1} \pi$. Hence according to the Lorenz gauge, we get $\p_r
A_r\big|_{r\rar \infty} = -\p_z A_z = 0$. As we discussed in the
introduction, on the IR boundary $A_r\big|_{z=z_m}=0$ due to the merging
of the Left and Right branes at this point (\ref{bc_IR}).\footnote{In
  Ref.~\cite{Domenech:2010aq} it was also noted that this condition is
  required by gauge invariance of the CS term in the ``hard-wall''
  model.} 
This does not apply though to the $z$ component and thus the gauge
condition will take the form $\p_z A_z\big|_{z=z_m} =-\p_r A_r
=0$. The boundary condition on the UV boundary $z=0$ is dictated by
the absence of the sources for the spatial axial current, hence
$A_r\big|_{z=0} = 0$ and $\p_z A_z = -\p_r A_r =0$. Finally in the
center of the soliton the pure gauge condition states
$A_z\big|_{r=0}=0$ and consequently $\p_r A_r\big|_{r=0} = 0$. The
boundary conditions for the Abelian vector 
field $v$ are set by the absence of sources, regularity at the center
and on the IR wall as well as the fall-off condition at infinity. 
Reexpressing these results in terms of the scalar fields 
$\eta_\alpha, \chi_\alpha$
we get the set of the boundary conditions shown in Table
\ref{BS_scalars}, which we will use for the numerical
solution. Conveniently, all of them are either of Dirichlet or of
Neumann type. 

\begin{table}[ht]
\begin{tabular}{|r@{\,=\,}l|r@{\,=\,}l|r@{\,=\,}l|r@{\,=\,}l|}
\hline
  \multicolumn{2}{|c|}{$r \rar \infty$} & \multicolumn{2}{c|}{$z=0$} & \multicolumn{2}{c|}{$r=0$} & \multicolumn{2}{c|}{$z=z_m$} \\ \hline 
 $\eta_1$ & $\cos \big(-\frac{\pi}{2} + \pi  \frac{z}{z_m}\big)$ &
  $\eta_1$ & $0$ & $\eta_1$ & $0$ & $\eta_1$ & $0$ \\
 $\eta_2$ & $\sin\big(-\frac{\pi}{2} + \pi  \frac{z}{z_m}\big)$ &
  $\eta_2$ & $-1$ & $\eta_2$ & $-1$ & $\p_z \eta_2$ & $0$ \\
 $\chi_1$ & $(m z + \sigma z^3) \cos\big(\pi  \frac{z}{z_m}\big)$ &
  $\chi_1$ & $(m z + \sigma z^3)$ & $\p_r \chi_1$ & $0$ & $\chi_1$ & $-(m z + \sigma z^3)$ \\
 $\chi_2$ & $(m z + \sigma z^3) \sin\big(\pi \frac{z}{z_m}\big)$ &
  $\chi_2$ & $0$ & $\chi_2$ & $0$ & $\chi_2$ & $0$ \\
 $\p_r A_r$ & $0$ & $A_r$ & $0$ & $\p_r A_r$ & $0$ & $A_r$ & $0$ \\
 $A_z$ & $z_m^{-1} \pi$ & $\p_z A_z$ & $0$ & $A_z$ & $0$ & $\p_z A_z$
  & $0$\\
 $v$ & $0$ & $v$ & $0$ & $\p_r v$ & $0$ & $\p_z v$ & $0$ \\
 \hline
\end{tabular}
\caption{\label{BS_scalars} Boundary conditions for the fields used in
  the numerical calculation.}
\end{table}

We look for a solution to the equations of motion, which follow from
Eqs.~(\ref{SYM}),(\ref{SU1}-\ref{SCS}). In order to reduce the problem to
the calculation on a square patch $[0,1]_x \times [0,1]_y$ we
introduce the rescaled coordinates 
\begin{equation}
\label{coords}
r = c \ \mathrm{tan}\Big( \frac{\pi}{2} x \Big), \qquad z = z_m \ y,
\end{equation}
where the dimensional constant $c$ specifies the radial scale on which
the solution will be best resolved. On this patch the homogeneous grid $N_x
\times N_y$ is introduced and we construct 7 equations (one for each
dynamical field) on each node by the nearest neighbor discretization
of the derivatives in the equations of motion. 

Special attention should be paid to the boundaries $r=0, z=0$ and 
$r \rar \infty$, as the equations of motion are divergent there. We do
not perform the regularization by stepping out from the
boundary. Instead, for the fields with Neumann boundary
conditions, we expand the equations of motion near the singular point
and solve the leading order contribution. This allows us to impose the
boundary conditions directly on the boundaries. 

After discretization we obtain $7 \times (N_x-2) \times (N_y-2)$
algebraic equations in internal points of the grid plus $4 N_x + 4
N_y$ equations on the boundaries and solve the resulting system both
by Newton-Raphson method\footnote{Strictly speaking we use the
  two-step procedure described in Ref.~\cite{Bolognesi:2013nja}: first we
  solve for all the fields except $v$, then we solve the linear
  equation for $v$ and so on. This is due to the inverse sign of the
  $v$ kinetic term, which prevents us from finding the solution as a
  simple minimum of the action functional.}\footnote{We use Wolfram
  Mathematica 9 \cite{Mathematica} for deriving the equations and
  compiling the numerical code and the \texttt {LinearSolve[]} procedure
  therein to invert the resulting matrix.} and the relaxation method,
independently, in order to have a good check of the numeric
results. The analysis of numerical accuracy is described in the
Appendix. 

As a starting field configuration we take the superposition of two
vortices according to Fig.~\ref{scheme}, which has unit topological
charge (\ref{baryon_charge}).  
At the end of the day we were able to find solutions with a number of
different parameters $\sigma$ and $m$ on the grids $60\times60$,
$120\times120$, $240\times240$ and observe good convergence and
matching between the two methods (several other grids were used as
well, see the Appendix). The procedure is run until the change in the
field values on each step falls below $10^{-8}$. For the solutions we
check that they do not depend on the value of the Lagrange multiplier
$\bm{\lambda}$, ensuring that the gauge fixing condition (\ref{gauge_fix})
is fulfilled. On top of that, we check that the observables do not
depend on the choice of the parameter $c$, which defines the radial
coordinate transformation (\ref{coords}).

\section{\label{sec:mass}Baryon mass}

\begin{figure}[ht]
\begin{tabular}{ccc}
\includegraphics[width=0.3\linewidth]{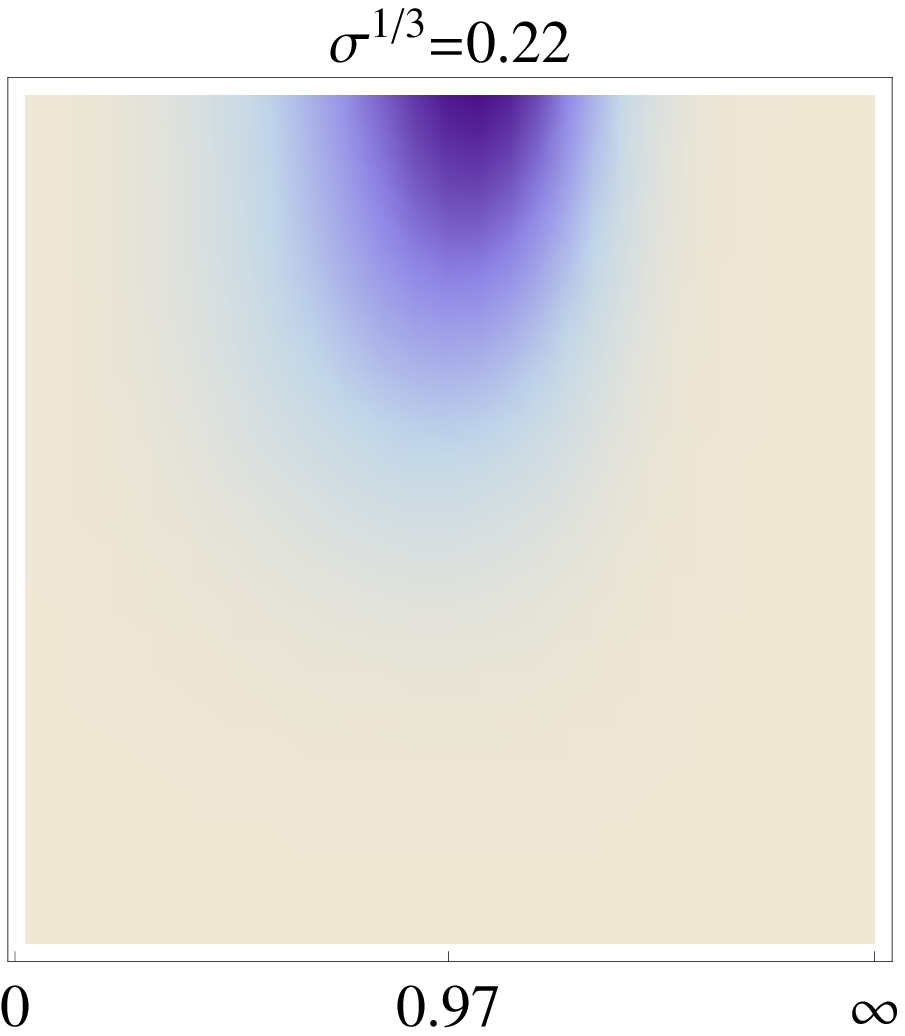} &
\includegraphics[width=0.3\linewidth]{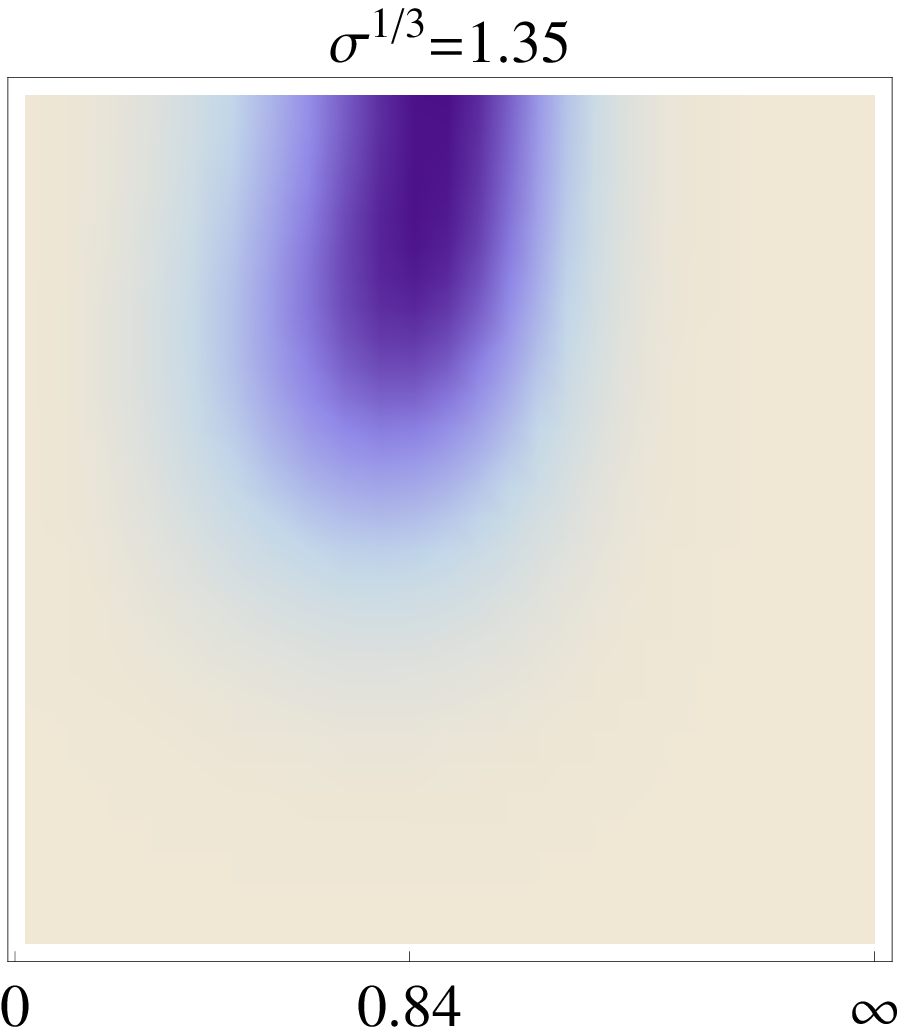} &
\includegraphics[width=0.3\linewidth]{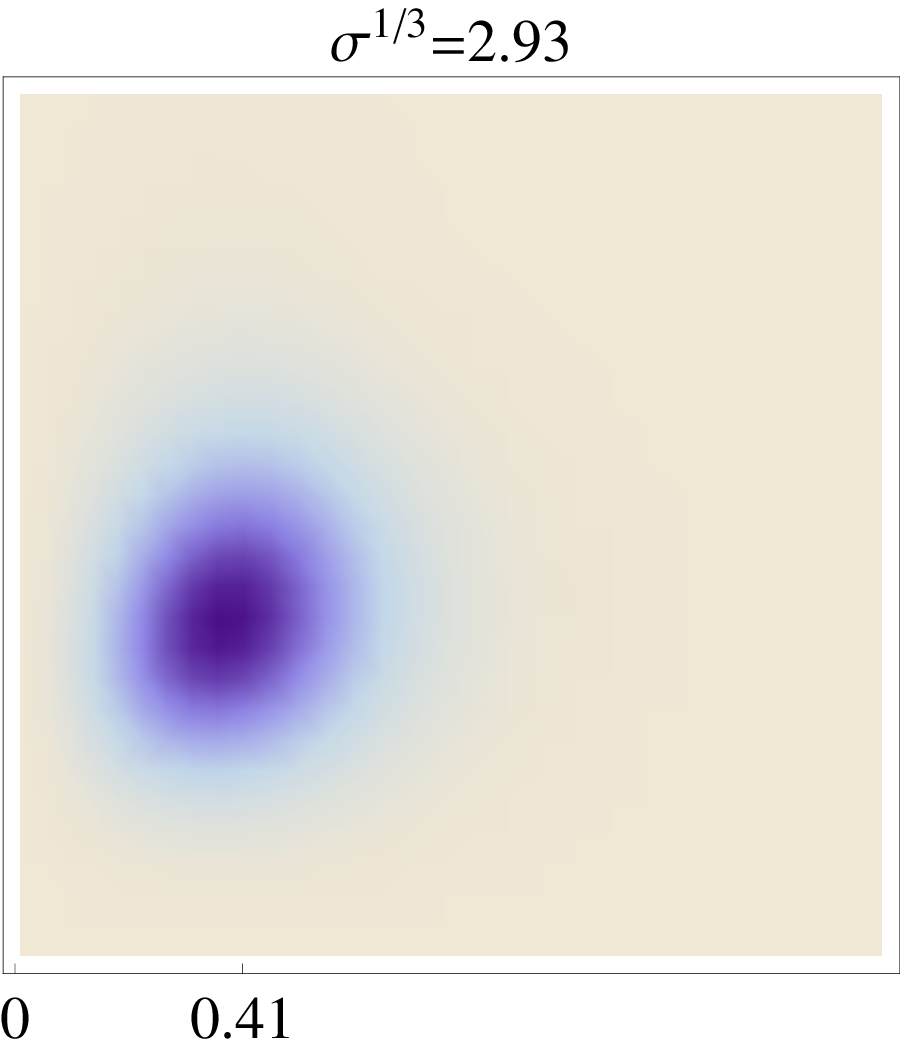} 
\end{tabular}
\caption{\label{tops}Topological charge density of the soliton at
  various values of the scalar field boundary value. At large $\sigma$ the
  soliton is detached from the IR wall. The radial (horizontal)
  coordinate and $\sigma$ are measured in units of $z_m=(323
  \ \mathrm{MeV})^{-1}$. } 
\end{figure}

Before we proceed with the analysis of various baryon parameters, it is
interesting to look at the general behavior of the solution. The
distribution of the topological charge at various $\sigma$ is shown in
Fig.~\ref{tops}. At very low $\sigma$ the solution is concentrated
near the IR boundary and does not feel the presence of the scalar
field. At intermediate values of the chiral condensate the charge
distribution is pulled downwards by the scalar ``repulsion force''
described above. Finally, at large $\sigma$ one can see that the solution
is detached completely from the IR wall and is localized at 
intermediate values of the holographic coordinate $z$. Another
observation is the 
dependence of the radial coordinate of the solution on the value of
the condensate. At large $\sigma$ the  
shrinking force produced by the scalar part of the action grows and it
is harder for the CS term to stabilize the radius of the solution. All
these observations are in perfect agreement with the intuition
developed in Sec.~\ref{sec:structure}. 

\begin{figure}[ht]
 \includegraphics[width=0.6\linewidth]{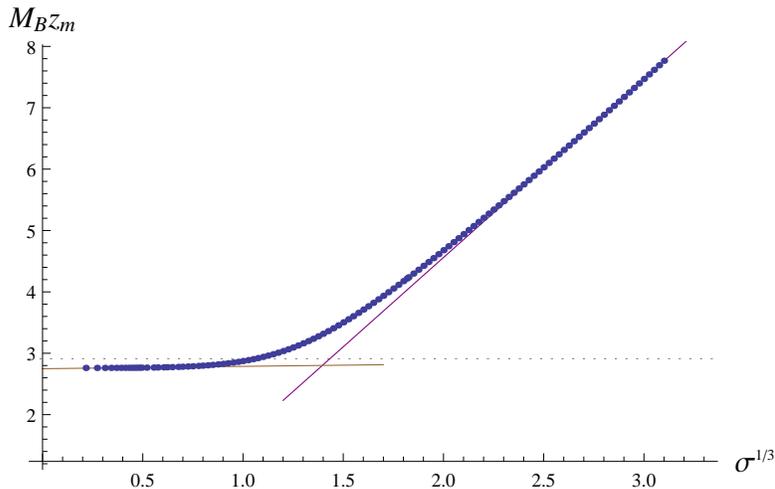}
\caption{\label{masses} Dependence of the baryon mass on the chiral
  condensate. The solid lines show the asymptotes at small and large
  condensate values. The dashed line shows the physical value of the
  proton mass.}  
\end{figure}

In order to study the mass of the holographic baryon we first
regularize the energy by subtracting the infinite contribution from
the vacuum configuration of the scalar field (\ref{vac_X}): 
\begin{equation*}
E_{reg} = \int dz d^3 x \left[ - \mathcal{L} + 2 \pi g_X^2 \frac{2 r^2 }{z^3} \left(m^2 - 3 \sigma^2 z^4 \right) \right]. 
\end{equation*}
The mass of the holographic baryon for different values of $\sigma$ is
shown in Fig.\,\ref{masses} (the value of quark mass here is $m=0.01
z_m^{-1} = 3.23 \, \mathrm{MeV}$ and we use $N_c=3$ in order to obtain the
numeric results). As anticipated, at low values of $\sigma$, when the
solution lies on the IR wall, the mass is governed by the confinement
scale. In the opposite case of large $\sigma$, the position of the solution
is only controlled by the scalar ``repulsion force'' and hence its
mass is directly proportional to the chiral symmetry-breaking energy
scale $\la \bar{q}q \ra^{1/3}$. The intermediate region, where the
effects of chiral and confinement scales are comparable, is quite
narrow, so using Eq.~(\ref{sigma}) we may approximate the result for the
mass of the baryon as  
\begin{align*}
M_B = N_c z_m^{-1} \, &\mathrm{max}\Big[ 0.92; -0.42 + 0.97 \ \left(2 \pi^2 \tfrac{\la \bar{q} q \ra}{N_c} \right)^{1/3}  \Big] \\
= \, &\mathrm{max}\Big[ 887; -407 + 940 \ \left(z_m \sigma^{1/3} \right) \Big] \mathrm{\ MeV}. 
\end{align*}
It is important to check that the $N_c$ scaling of the baryon mass is
linear due to $\la \bar{q}q \ra \sim N_c$. 
Interestingly, the physical value of the baryon mass $M_B=940\,\mathrm{MeV}$
corresponds to the point where the scalar repulsion just becomes
significant (as claimed in \cite{gkk}).  

\section{\label{sec:form}Mean values of quark operators}

Now we calculate the mean values of the currents induced by the
baryon. The vacuum expectation value of the operator is obtained by
taking the variation of the action on the classical solution with
respect to the boundary value of the corresponding field
\cite{gubser-klebanov, Bianchi:2001kw}. Given the field content of the
model, we are able to calculate the mean values of vector and axial
quark currents as well as the scalar quark bilinear. It is important
to note also that the proper formulation of the operator/field duality
is possible only in a specific gauge in the 5D theory, namely in the axial gauge
$A_z = 0$. Our numerical solution was obtained in different, Lorentz
gauge (\ref{gauge_fix}), hence a gauge transformation should be
performed before we can obtain the mean values. Let us denote the fields in
the axial gauge with bars: $\bar{A}_z = 0$. The action (\ref{SYM})
evaluated on the classical solution in axial gauge can be recast as a
boundary term\footnote{The contribution from $z=z_m$ is canceled by
  an appropriate boundary term on the IR wall and the contribution
  from the CS term vanishes because of the zero boundary value of the
  Abelian vector field $v$.} 
\begin{equation*}
S_{cl} = \int dt dr  \ 4 \pi r^2 \ \left(-\frac{1}{g_5^2}\right)  \bigg\{ 
\frac{1}{z} \bar{A}_r \p_z \bar{A}_r 
+ \frac{2}{r^2 z} \left[ \bar{\eta}_2 \p_z \bar{\eta}_2  +  \bar{\eta}_1 \p_z \bar{\eta}_1 \right]
+ \frac{3}{z^3}  \left[
   \bar{\chi}_1 \p_z \bar{\chi}_1  + \bar{\chi}_2 \p_z \bar{\chi}_2  \right]  
\bigg\}_{z = \epsilon}.
\end{equation*}
According to the equations of motion, the solutions near the boundary
are represented by the Frobenius series 
\begin{equation}
\label{frobenius}
\bar{A}_r = a^{(0)} + a^{(2)} z^2 + \dots, \qquad  \bar{\eta}_\alpha = \eta^{(0)}_\alpha + \eta^{(2)}_\alpha z^2 + \dots, \qquad \bar{\chi}_\alpha = \chi^{(1)}_\alpha z + \chi^{(3)}_\alpha z^3 + \dots.
\end{equation}
Substituting these into the action we find a divergent contribution as $\epsilon \rar 0$
which comes from the $\chi$ term: 
$S_{div} \sim \epsilon^{-2} \big(\chi_\alpha^{(1)} \big)^2$. 
This must be regularized according to the
holographic regularization procedure \cite{Bianchi:2001kw} with a
boundary counter term 
\begin{equation}
S_{c.t.} = \int dt dr \ 4\pi r^2 \left( \frac{3}{g_5^2} \right) \frac{1}{\epsilon^4} \left[\bar{\chi}_1(\epsilon)^2 + \bar{\chi}_2(\epsilon)^2 \right]. 
\end{equation}
In order to get the mean value of the 3D spatial current, one needs to
take the variation of the regularized action 
$S_{reg} = S_{cl} + S_{c.t.}$ with respect to the boundary value of
the corresponding spatial 
component of the 3D vector field $A^{(0)}_i$ or $V^{(0)}_i$. It is
convenient to express these boundary values in terms of the 2D fields
of the ansatz (\ref{ansatz}). This will reduce the problem to the
variations with respect to $\eta_\alpha^{(0)}, \chi_\alpha^{(1)}$ and
$a^{(0)}$.  
Therefore in terms of the Frobenius coefficients (\ref{frobenius}) we
get\footnote{Note that if one substitutes the vacuum solution
  (\ref{vac_X}) into the expression for $\la\bar{q}q\ra$ one gets the
  relation (\ref{sigma}) between $\sigma$ and $\la\bar{q}q\ra$.} 
\begin{align}
\label{VEVs}
\la J^V \ra^a_i &= - \frac{2}{g_5^2} \frac{\eta_2^{(2)}}{r} \ \epsilon_{i a k} \frac{x_k}{r}, \\
\notag
\la J^A \ra^a_i &= - \frac{2}{g_5^2} \left[ \frac{\eta_1^{(2)}}{r} \ \left(\delta_{ia} - \frac{x_i x_a}{r^2} \right) + a^{(2)} \ \frac{x_i x_a}{r^2} \right], \\
\notag
\la \bar{q} q \ra_{\rho \gamma} &= - \frac{6}{g_5^2} \left[ \chi_1^{(3)} \ \frac{\mathbbm{1}}{2}  + i \chi_2^{(3)} \ \frac{x^a t^a}{r} \right]_{\rho \gamma}.
\end{align}

The values of the coefficients can be calculated after we transform
the numerical solution to the axial gauge. With the gauge function
$\lambda$, the fields in Eq.~(\ref{SYM}) transform as 
\begin{equation}
\bar{A} = A + \p \lambda, \qquad 
\bar{\eta}_\alpha = \eta_\alpha \cos \lambda - \epsilon^{\alpha \beta} \eta_\beta \sin \lambda , \qquad
\bar{\chi}_\alpha = \chi_\alpha \cos \lambda - \epsilon^{\alpha \beta} \chi_\beta \sin \lambda. 
\end{equation}
In order to reach the axial gauge $\bar{A}_z =0$ while keeping
the value of the quark mass $m \sim \chi_1^{(1)}$ constant, we choose  
\begin{equation}
\lambda = - \int_{0}^{z_m} dz A_z. 
\end{equation}
In the vicinity of the boundary, it is expanded as $\lambda(\epsilon)
= - \epsilon A_z(\epsilon)$. 
At the end of the day the coefficients in Eq.~(\ref{VEVs}) are expressed
in terms of the components of our numerical solution 
\begin{align}
\label{num_coeffs}
\chi_\alpha^{(3)} &= \left. \frac{1}{2 z} \p_z \frac{\bar{\chi}_\alpha}{z} \right|_{z = \epsilon} = \frac{1}{2}  \left[ \frac{1}{z} \mathcal{D}_z \Big(\frac{\chi_\alpha}{z} \Big) + \epsilon^{\alpha \beta} A_z \, \mathcal{D}_z \Big(\frac{\chi_\beta}{z} \Big) \right]_{z=\epsilon}, \\
\notag
\eta_\alpha^{(2)} &= \left. \frac{1}{2 z} \p_z \bar{\eta}_\alpha \right|_{z = \epsilon} = \frac{1}{2} \left[ \frac{1}{z} \mathcal{D}_z \eta_\alpha + \epsilon^{\alpha \beta} A_z \, \mathcal{D}_z \eta_\beta \right]_{z=\epsilon}, \\
\notag
a^{(2)}&= \left. \frac{1}{2 z} \p_z \bar{A}_r \right|_{z = \epsilon} = \frac{1}{2} \left[ \frac{1}{z} (\p_z A_r - \p_r A_z) \right]_{z=\epsilon}. 
\end{align}
Given these expressions we can calculate the mean values of the currents (\ref{VEVs}). 

\begin{figure}[!ht]
\includegraphics[width=0.45 \linewidth]{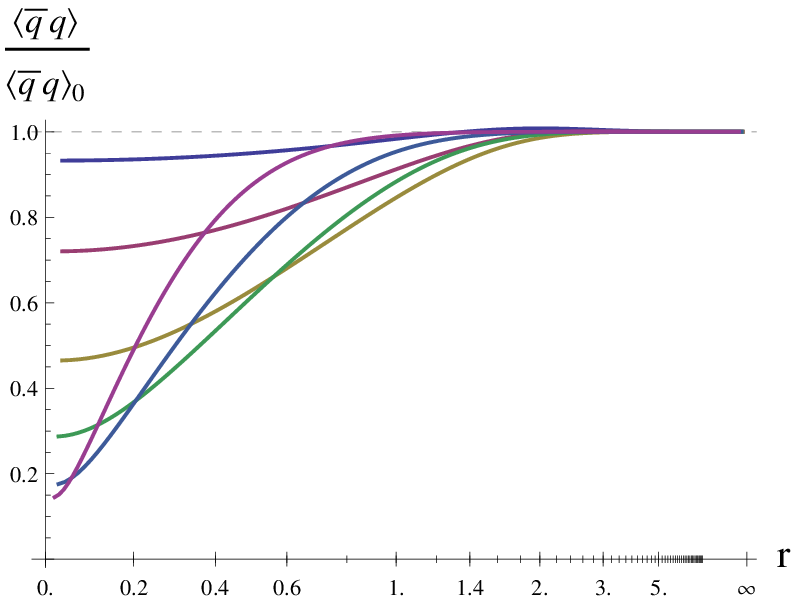} 
\includegraphics[width=0.45 \linewidth]{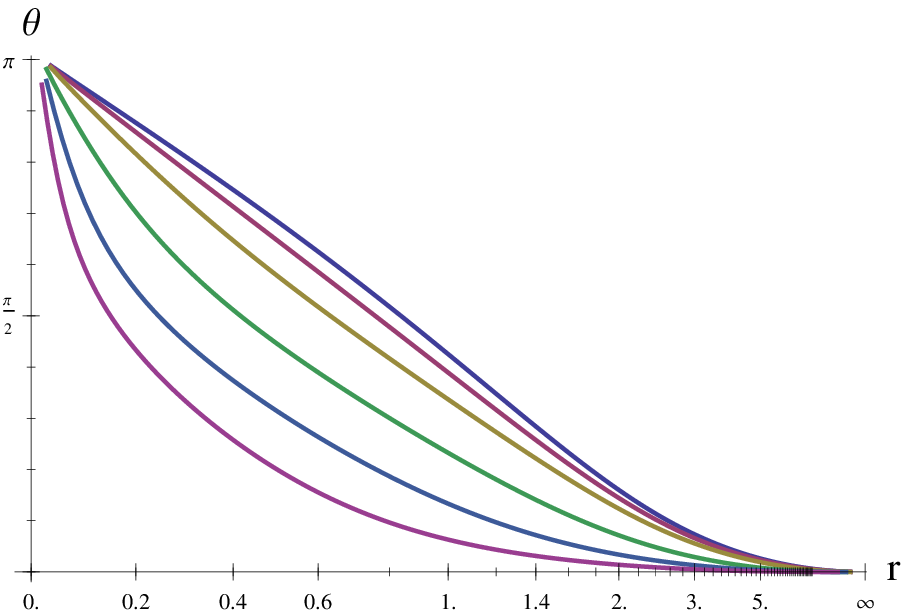} 
\caption{\label{condensates} Relative local mean value of the scalar operator $\la \bar{q} q \ra$  and its  phase $\theta$ for various solutions \hbox{with $\sigma^{1/3} z_m =\{0.31, \ 0.37, \ 0.62, \ 1.50, \ 2.10, \ 3.10\}$} counting from top to bottom.}
\end{figure}

It is interesting to study the local scalar VEV. In a hedgehog
Skyrmion it behaves as 
\begin{equation}
\la \bar{q} q \ra (r) = \la \bar{q} q \ra_0 \left(\cos(\theta) + i \frac{t^a x^a}{r} \sin(\theta) \right), 
\end{equation}
where $\la \bar{q} q \ra_0$ is the value of the chiral condensate at
infinity, and $\theta(r)$ is a chiral phase which goes from $\pi$ in
the center of the Skyrmion to 0 at infinity. In the analysis of our
solution we find that the phase of the scalar VEV behaves exactly as
expected (see Fig.~\ref{condensates}), but on top of that, the modulus
of the chiral condensate is suppressed in the center of the
baryon. This observation agrees with the lattice studies
\cite{Iritani:2014fga} of the partial chiral symmetry restoration
inside baryons. Spectacularly, while the value of the chiral
condensate affects the baryon mass, the presence of thee baryon has its
effect on the scalar quark density as well. 

The other interesting observable is the nucleon axial charge $g_A$. In
Refs.~\cite{Panico:2008it, Domenech:2010aq} the value of the axial charge
calculated for a holographic baryon was almost 2 times smaller than
the phenomenological value $g_A
\sim 1.25$, although the other baryon observables
exhibited good agreement. One should expect, though, the dependence of
$g_A$ on the pion mass, which is related to the chiral condensate by
the Gell-Mann-Oakes-Renner relation. Thus it is interesting to check
whether this value changes significantly for different values of
$\sigma$. For this purpose we reproduce the calculation of
Ref.~\cite{Domenech:2010aq} and study the dependence of $g_A$ on the chiral
condensate. If one parametrize the mean value of the QCD axial quark current by the following radial functions 
\begin{align}
 \la J^A \ra_i^a (\mathbf{r}) &= A_1(r) \delta^{i a} + A_2(r) \frac{x^i x^a}{r^2}, 
\end{align}
the axial form factor is expressed as \cite{mei}
\begin{equation*}
 G_A (q^2) = - \left(\frac{M_B}{E} \right) \frac{8 \pi}{3} \int \limits_0^\infty \left[ r^2 j_0(qr) A_1 (r) + \Big(\frac{r}{q}\Big) j_1(qr) A_2(r) \right],
\end{equation*}
and the axial vector charge $g_A$ is
\begin{align}
 \label{axial_formfactors}
 g_A &= G_A(\mathbf{q}^2=0) = - \frac{8 \pi}{3} \int \limits_0^\infty dr \, r^2 \left[ A_1(r) + \tfrac{1}{3} A_2(r) \right].
 \end{align}
The values of $g_A$ for different $\sigma$s and different quark masses
$m$ are shown in Fig.~\ref{gAs} (for details of the numerical
calculation see the Appendix). 
At $\sigma^{1/3}=1.23$ (which provides a good fit to QCD
phenomenology) we get 
$ g_A = 0.63 $, which is close to the value obtained in
Ref.~\cite{Domenech:2010aq}. The axial charge exhibits only a minor
dependence on the parameters of chiral symmetry breaking. So we can conclude that the
discrepancy with phenomenological value is a qualitative feature and can not be cured by a simple tuning of the model. We note
that this result is similar to the Skyrme model, where $g_5$ also has 
a small value \cite{Adkins:1983ya}, which is weakly affected by the pion mass. The problem of small $g_A$ is a long standing problem in Skyrmion physics. There might be significant $N_c$ corrections to this value, which are not taken into account here and may require a qualitative modification of the model. For a detailed discussion of this issue see e.g. \cite{Panico:2008it, mei} and references therein.  

\begin{figure}[!ht]
\includegraphics[width=0.6 \linewidth]{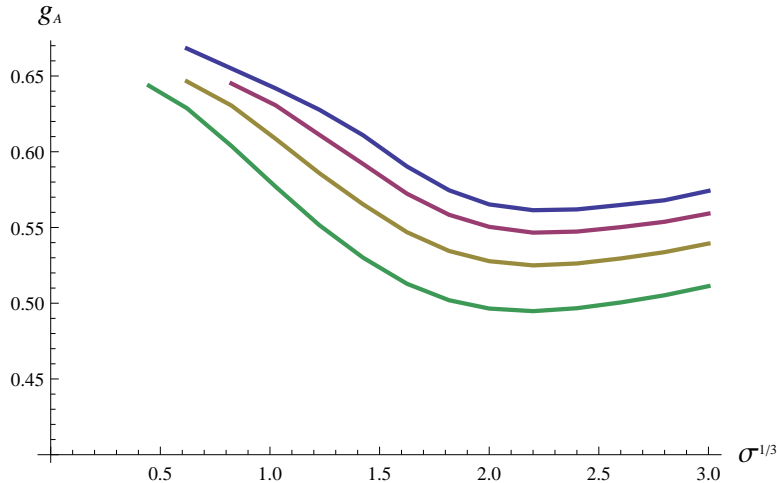} 
\caption{\label{gAs} The baryon axial charge for various $\sigma$s and various quark masses: from top to bottom \hbox{$m  z_m = \{ 0.01, 0.02, 0.04, 0.08 \}$} ($z_m^{-1} = 323\,\mathrm{MeV}$).}
\end{figure}

\section{\label{sec:conclusion}Conclusion}

In this paper we have studied the interplay between chiral symmetry
breaking and the baryon in a holographic QCD setup. Although the numerical
calculations were done for a specific ``hard-wall'' model, our
qualitative treatment is fairly general and the results should hold
for a generic model. We have proved in a clear-cut manner the
conjecture advocated in Refs.~\cite{gk,gkk} that there are essentially two
regimes in baryon physics where the dependence of the baryon mass
on the chiral condensate 
is very different. The numerical calculations show that the physical
value of the baryon mass is in the transition region between the two
regimes, which clearly demonstrates that the Ioffe's formula for the
baryon mass \cite{ioffe} derived in QCD sum rules is relevant. 

The main finding of our work is the phenomenon of repulsion of the
baryon from the IR wall due to the interaction with the scalar field
dual to the chiral condensate. It is very important to understand how
this repulsion looks from the perspective of the Sakai-Sugimoto
setup. At large values of the chiral condensate the baryon is moved away
from the tip of the cigar. The only way to do this without violating
P-symmetry is to \textit{split} the soliton into two symmetric
halves. One of them is moved to the Left brane and another to the
Right one. For symmetry reasons, these two halves carry equal
half-integer baryon charges and opposite half-integer Abelian axial
charges, producing in total the solution with unit baryon charge.

In spirit the splitting of the holographic baryon is very similar
to the splitting of an instanton on compact space into monopoles. This
phenomenon takes place when the system is considered at finite
temperature or finite chemical potential. However, the details can be
very different and deserve a thorough study. In particular it is not
completely clear how the 
splitting occurs in terms of the branes.  The baryon is
represented by the D4 brane wrapped around the internal sphere $S^4$ 
and cannot be naively split into half D4 branes. The process
could be viewed upon a T-duality transformation around the 
compact coordinate on the cigar. Another delicate issue concerns the
behavior at the nonvanishing temperature. Different scenarios of
the splitting phenomenon could take place once one takes into account
the temperature dependence of the chiral condensate. 

We should recall, that once one moves from the tip of the cigar,
the separation between $D8$ and $\overline{D8}$ branes grows with the radius of
the compact circle $\tau$. In our treatment we neglected this separation
in favor of the opportunity of working with the 5D model with local
fields, but in a general setup once the baryon is split and repelled
from the tip of the cigar this separation may lead to interesting
phenomena. For example, although the net Abelian axial charge of the
solution is zero, the axial dipole moment is nonzero and
proportional to the size of the $\tau$-circle.  The $\theta$-
dependence of the Skyrmion due to the splitting can be considered in the SS model as well since the $\theta$-term corresponds to the holonomy of the
one-form RR field along the compact coordinate on the cigar. One
should note though, that the splitting effect is proportional to the
size of the instanton and therefore is parametrically suppressed by
the 't Hooft coupling in the original Sakai-Sugimoto model. On the
other hand, the divergent tachyonic mode of the open string can
provide a sufficiently large scalar boundary value, which will render the effect sizable. Given these subtleties, it is not clear in which
particular setup one should study the phenomena mentioned above. 

The other interesting finding, that we made in the present study, is
the backreaction of the baryon solution on the local density of the
quark bilinear operator. We find that the chiral condensate gets
suppressed inside the baryon and the chiral symmetry is partially
restored. This idea was widely discussed in the literature but to our
knowledge this phenomenon has not been directly observed in
any models of 
QCD.  The chiral condensate in the SS model follows from tachyon
condensation 
on the string, hence it would be interesting to investigate the impact
of the Skyrmion on the tachyon action directly. This could also be 
related to the recent discussion of the spectrum of hadrons including
baryons in terms of the details of the Dirac operator spectrum in QCD
\cite{glozman}.

\acknowledgments
The authors acknowledge the contribution of Peter Kopnin during the early stages of this project. 
We benefited from the discussions with Paul Sutcliffe, Daniel Arean and Nick Evans.
A.K. is grateful to the Mainz Institute for Theoretical Physics (MITP) for its hospitality and its partial support during the completion of this work.

The work of A.G. and A.K. is partially supported by RFBR grant 15-02-02092.
\appendix*
\section{\label{Accuracy} Accuracy of the numerical calculation}
\begin{figure}[!ht]
\includegraphics[width=0.6 \linewidth]{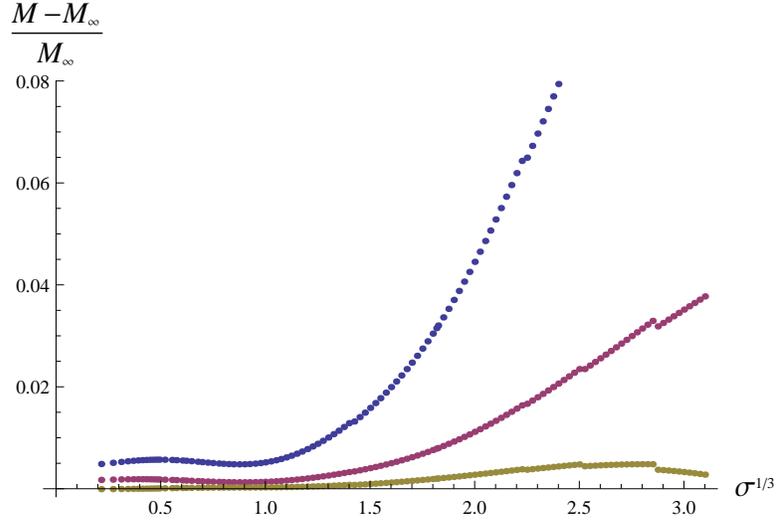} 
\caption{\label{prec_m} The accuracy of calculated baryon mass on
  different $N\times N$ grids with $N=\{60, 120,240\}$ (from top to
  bottom). $M_\infty$ is the extrapolated value for $N \rar \infty$.}
\end{figure}

We perform our numerical analysis using two different methods
implemented in different codes in order to exclude any possible
systematic errors (see text). The mass and topological charge
distribution of the soliton is calculated on the grids $60\times60$,
$120\times120$ and $240\times240$. One can observe a good convergence
of the results (see Fig.~\ref{prec_m}) and the estimated precision of
the finest grid is within 1\%.  

\begin{figure}[!ht]
\includegraphics[width=0.6 \linewidth]{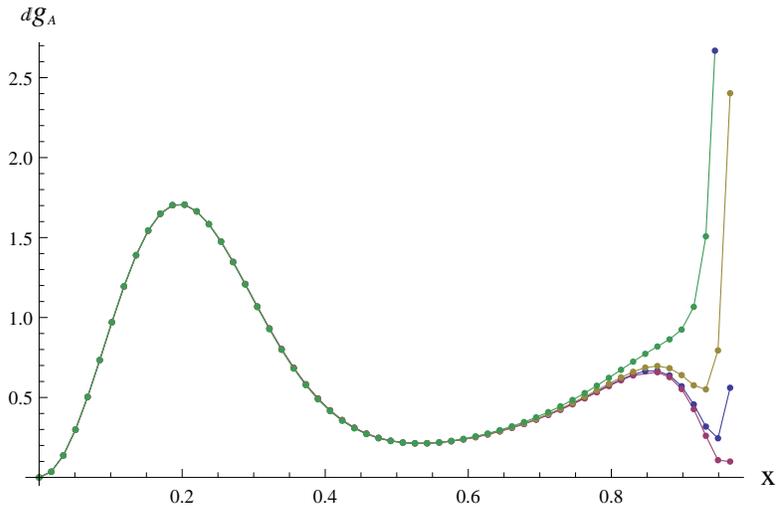} 
\caption{\label{prec_dg} The integrand (\ref{axial_formfactors})
  calculated for $\sigma=2.6$ on different grids with $N_y = \{120,
  240, 480, 1000 \} $ (from top to bottom). The tail is decently
  resolved only on the finest grid.} 
\end{figure}

It is numerically challenging to calculate the axial charge
(\ref{axial_formfactors}). The mean value of the axial current is
defined by the asymptotic form of the solution near the boundary $y = \epsilon
\rar 0$. The equations of motion are singular on this boundary,
hence the precision of the calculated solution should be higher than
$\epsilon$. On top of that, the integrand in Eq.~(\ref{axial_formfactors})
has several factors of $r$ and in order to get finite results one needs
to ensure that the fall off of the solution at large $r$ is calculated
with a precision of order $1 / r^2$. An additional difficulty is the
fact that we impose nontrivial boundary conditions at $r \rar \infty$. 
Because of that, the accuracy of the solution at $r\rar \infty$ is 
limited by the precision of the finite difference approximation
of the scalar field derivatives on the grid. At the end of the day we are
forced to use very dense grids in the $z$ direction in order to obtain
convergent integrals for $g_A$. The sample of the integrand in
Eq.~(\ref{axial_formfactors}) for $\sigma = 2.6$ and  $m=0.01$ calculated
on the grids $60 \times 120$, $60 \times 240$, $60 \times 480$ and $60
\times 1000$ is shown in Fig.~\ref{prec_dg}. While for intermediate
values of $x$, the results of all the grids coincide, the tail at $x=1$ ($r\rar \infty$) is
resolved only on the finest one.  

\end{document}